# Sound masking degrades perception of self-location during stepping: A case for sound-transparent spacesuits for Mars

Jose Berengueres, Maryam Al Kuwaiti, Ahmed Yasir and Kenjiro Tadakuma

*Abstract*— Most efforts to improve spacesuits have been directed towards adding haptic feedback. However, sound transparency can also improve situational awareness at a relatively low cost. The extent of the improvement is unknown. We use the Fukuda-Unterberger stepping test to measure the accuracy of one's perception of self-location. We compare accuracy outcomes in two scenarios: one where hearing is impaired with sound masking with white noise and one where it is not. These scenarios are acoustic proxies for a sound muffling space suit and a sound transparent space suit respectively. The results show that when sound masking is applied, the error in self-location increases by 14.5cm, 95% CI [4.04 28.22]. Suggestions to apply the findings to Mars spacesuit designs are discussed. A cost-benefit analysis is also provided.

*Clinical Relevance*— A paired t-Test for unequal variances applied to the drift distance outcome of the Fukuda-Unterberger stepping test indicates that sound masking with white noise increases the subjects' self-location error by 14.5cm, 95% CI [4.04 28.22].

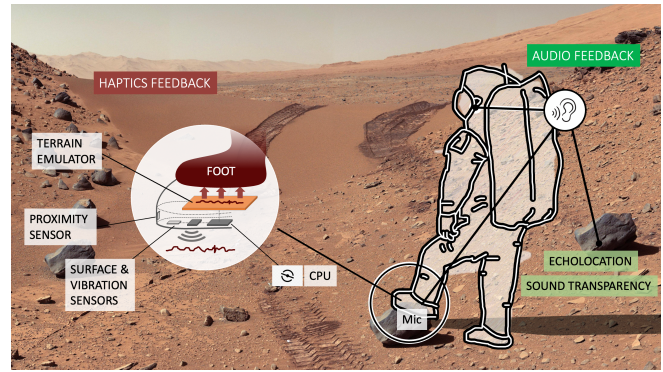

Figure 1. A sensory enhanced spacesuit concept. Most efforts to improve suits have focused on haptics (left). However, sound transparency can also be used on Mars (right). Background: Mast Camera (Mastcam), Curiosity's (9 February 2014), NASA/Caltech.

## I. INTRODUCTION

### A. Suits impair cognition

Extra-Vehicular Activity (EVA) spacesuits, due to their thickness that ranges from 10 to 30 mm, and having up to 16 layers, impair the astronauts' dexterity and senses. Namely, (i) grasping dexterity, (ii) reduced range of motion, (iii) proprioception, and (iv) cognitive ability due to sensory deprivation. This degradation of capabilities can affect mission safety and increase, for instance, the risk of falls. For example, the Apollo lunar missions recorded 27 falls and 21 near falls according to [1]. The falls were attributed to the Lunar low gravity and Apollo spacesuit design which had a decreased range of motion and changed the center of mass due to the top-heavy portable life support system. The **inability to read the terrain** was cited as the main cause in the Apollo 15 analysis of astronauts' falls on the Moon in the report by Kubis [2]. More recently, new research has shown that spacesuit design is improvable. For example, [3, 4] found that suit designs have mismatched ankle kinematics and inadequate contact between the foot and boot. [5] analyzed all EVA activities between 1998 and 2001 and identified similar issues related to spacesuit boot fit. Further, it was determined that such issues could make spacewalks a difficult task for astronauts due to contact and fatigue injuries [4]. Recent suit designs such as NASA's Artemis program are an improvement compared to the 1960-80s suits [5]. However, they still impair the human senses such as hearing, the field of view, (astronauts can barely see their own feet), proprioception, and grasping dexterity.

Moreover, Moon suits like Artemis have the extra design constraint that they should not collect the toxic Moondust [5]. In summary, on one hand, the extraordinary physical demands placed on EVA suits result in degraded sensory input to the astronaut which decreases their cognitive ability [6,7]. On the other hand, this is a known factor in higher accident rates [8]. These interdependent constraints make suit design challenging.

### B. Improvement strategies

The two straightforward paradigms for improving EVA suits are: (i) better **mobility** as in [3-5], (ii) reducing the **sensory impairment** [6-8]. Hence, various authors proposed solutions that address (i), (ii), or a combination of both. By function, we can group these solutions in two groups: (a) **nimbler suits**, for example, as proposed in the compressing suit design of [9], and (b) making the suit more *transparent* to sensory inputs.

#### 1) Haptics

One popular way of implementing transparency is by way of haptics. In 2001, [10] designed a sole morphing astronaut boot with the purpose of letting the astronaut feel the ground and avoid falls. In 2017, [11] highlighted that future space-footwear technology should allow for comfortable and secure restraint. The authors concluded that such technology should adopt a **holistic approach** to human-centered design and the interrelated occupational health and safety in space. In 2018, [12] proposed design specifications for a spacesuit able to record the dynamics and kinematics of human-spacesuit interaction. [13] analyzed the influence of microgravity on

J.B., M. K., A.Y. are with UAE University, College of IT, Al Ain, AD, 17551, UAE (phone: +971-713-03-5555 fax: +971-713-03-5555; e-mail: {jose, 201900648, 201306379 }@ uaeu.ac.ae), K.T, is with Tohoku University, Sendai, Japan (tadakuma@rm.is.tohoku.ac.jp)

task performance. In 2021, [14] provided evidence that sensorimotor performance changes under microgravity conditions during a manual tracking task requiring very precise and continuous motions changes. Outside the aerospace field, various authors have proposed haptics to reduce fall risk [15], and improve posture [16], and numerous and varied biofeedback applications Commercially, various companies such as *HaptX*, have developed haptics for VR gaming that could be adapted to EVA suits. However, they are considered bulky and heavy. For example, HaptX, (an experimental high-resolution haptic glove), requires a 5kg backpack.

*2) Sound*

As we saw, the bulk of spacesuit research has focused on addressing sensory impairment via haptics or kinematics. Little attention has been paid to the potential of audio. While an explanation for this might be that there is no atmosphere to transmit sound on the Moon or the International Space Station (ISS), Mars does have an atmosphere. This presents us with the opportunity to apply *sound design* to spacesuits. The benefits of sound design are well known in various fields. In VR, for instance, there are several studies on the application of sound. [17] studied how to enhance walking sensations with sound, [18] how to implement body-centered auditory VR. In the field of human factors, studies have studied extensively audio alerts in-flight cockpits [19], cars, and so on. Note that while in outer space and the Moon there is no sound conducting atmosphere, that does not mean that the suit cannot improve with sound design. For example, during an EVA on ISS, artificially generated ambient sound via a *digital twin* of the surroundings could help the astronauts orient themselves more naturally by way of enabling the human echolocation sense, while at the same time mitigating sensory deprivation effects.

*C. Research questions*

Therefore, the first research question is, does adding sound transparency to an EVA suit improve proprioception? The second question is, by how much? Following, we report the results of the Fukuda-Utenberger stepping test [20].

## II. METHODS

*A. Design*

We chose the widely used Fukuda-Utenberger stepping test for its simplicity. The common use case of the test is to diagnose if the vestibular system is functioning properly. A list of environmental factors that affect the outcomes of this test in pathology-free participants was compiled by [21]. To test the hypothesis that a sound transparent suit improves proprioception (self-location), we set up two experimental settings. In **scenario A,** we test how a spacesuit (that muffles external sounds) impairs self-location. As we do not have access to an actual spacesuit, we have simulated the muffling effect by using sound masking [22]. We use white noise at 50% volume, a smartphone, and commercial headsets. The rest of the setup is as described in [20]. The participants are told to steps on the same spot 50 times with the headsets on and eyes closed. The goal is to minimize drifting. Afterward, we measure the distance traveled (**drift**), and the rotation of the body over its own vertical axis is measured. In **scenario B**, we test a hypothetical spacesuit that is completely sound transparent. To simulate this condition, the standard test [20] without any manipulation is administered. To avoid learning bias favoring A, subjects perform A, then rest for 5 minutes, then B.

*B. Measures*

We measure two outcomes. The drift in **rotation** and the drift in **distance**. A rotation of more than 45 degrees or a distance drift of more than 1m is considered pathological [20]. The start and end positions of the foot were photographed for record and are available in the annex. Due to covid, the campus was closed to students. Therefore, experiments were carried out by students at their respective locations in a decentralized manner.

*C. Participants*

14 Participants were recruited using word of mouth and slack channels between December 2021 and January 2022. The average age of participants was 33.6 years (SD 11.8). Inclusion criteria: (i) 18–65 years of age, (ii) free of foot or leg ailments, injury, or impairments. (iii) free of any sense of balance impairment or pathology such as vertigo. Of the initial intake, 6 were discarded for age reasons. Two were excluded because they suffered vertigo, (their drift exceeded 45 degrees and 1m well before the completion of the 50 steps). Of the remaining 6 participants, 3 were female. Half of the participants were employed full-time, the rest were students. Before the start, participants were provided a written description of the study and signed the consent form. Then received oral instructions. This study has been approved by the UAEU Social sciences ethics board.

## III. RESULTS

Fig. 2 shows a centroid chart of the distance drifts of the participants on the XY plane. It compares scenario A with scenario B. (See Annex for notebook code). The large symbols in the middle of each cluster are the means of each group. Table 1 shows the results of a paired t-Test for unequal variances that compares the distance drift of the two scenarios.

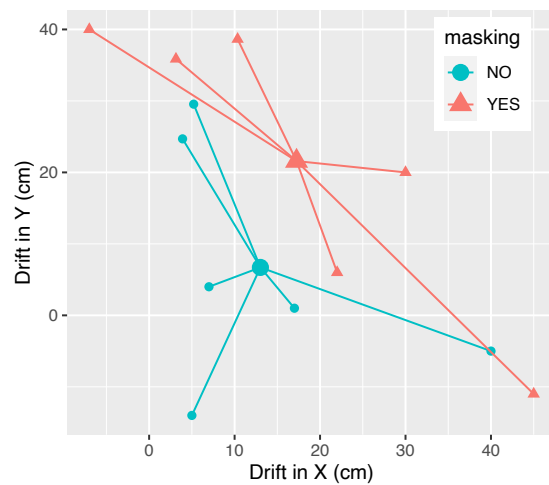

Figure 2. Centroids plot of the experimental data (N=6). Subjects start test at (0,0), looking towards the Y-axis positive direction. Masking of environment sounds increased the drift (△).

Table 2 shows additional CI statistics. An increase in distance drift 14.5 cm, 95% CI [4.04 24.04] cm was measured when sound masking was used. No significant drift in rotation was detected.

TABLE I. PAIRED T-TEST UNEQUAL VARIANCES

Distance from start point after 50 steps
(Fukuda-Utenberger stepping test)

|  | Scenario A  *Sound masking (muffling suit)* | Scenario B  *Normal hearing (transparent suit)* |
|---|---|---|
| Mean error distance (cm) | **37.0** | **22.5** |
| Variance of distance | 62.5 | 135.3 |
| Observations | 6.0 | 6.0 |
| t Stat | 2.5 |  |
| P(T<=t) one-tail | 0.017 |  |

TABLE II. STATISTICS FOR DIFFERENCES (A-B)

| Mean distance difference A-B | 14.4 |
|---|---|
| Standard Error | 4.0 |
| Median | 10.5 |
| Standard Deviation | 9.9 |
| Sample Variance | 97.8 |
| Confidence Level(95.0%) | 10.4 |
| Lower CI - Upper CI | [4.04 24.80] |

## IV. DISCUSSION

### A. Interpretation

We have measured how masking ambient sound degrades the accuracy of the perception of self-location when eyes are closed. The test is based on the principle that when visual input is not available we rely on the alternative inputs (of lesser quality) for orientation and balance (sensor fusion). An increase in the drift distance was detected. However, no significant drift in body orientation (rotation) was detected. We explain this because the vestibular system (our internal gyroscope) is not adept at detecting slow linear change. Nevertheless, curved path drifts are observed in studies where the vestibular system is strongly interfered with [23]. Note that earphone usage has been associated with various ear disorders [24].

### B. Applicability of findings

Current suit designs limit the field of view and ambient sounds like in scenario B. In particular, key visual cues from the ground are often blocked by the bulky suit, (the astronaut has no direct line of sight with their own feet). The data support the hypothesis that in such situations, providing sound input might reduce the risk of trip over due to restored echolocation [25] and mitigation of sensory deprivation [1,2,6,7]. Following we elaborate on the specifics of a sound transparent EVA for Mars.

### C. Implementing transparency on Mars

The emulation by the suit of the ambient sounds as they sound on Earth will likely result in improved cognition as an increase in cognitive load has been observed when sensory inputs are impaired [7]. However, the implementation is not straightforward. To start with, sounds on Mars require signal amplification. Eq. 1 is the sound Intensity equation (W/m$^2$).

$$I = p^2/\rho c \qquad (1)$$

on **Mars**,

$\rho$ = density air (0.02kg/m$^3$) 1.6% of Earth,
$c$ = speed of sound (240m/s) 70% of Earth,

therefore, an event that on Earth is heard with Intensity I, is heard on Mars with intensity 85 times lower. In other words, about 20dB of amplification are required for the suit of Fig. 1 to emulate Earth-like audio intensity.

### D. Benefits of a ground microphone

However, this required amplification is less if the microphone is located near the ground (in the boot) instead of the helmet. This boot location is of particular interest, as the thumping of the foot on the terrain is the main source of sounds we are interested in proprioception. The thumping of each step, not only helps the subjects echolocate their own feet, but the echo can also inform about the distance, size, and nature of nearby objects as proven by [25]. In addition, the **sound** of a boot when landing at each step provides useful sound (and haptic) cues about the terrain (muddy, sandy, soft, crunchy…). The Signal Noise Ratio of these sounds is enhanced when the microphone is placed on the boots. From field tests, (not reported here for lack of space), subjects reported that hearing the sound from microphones placed on boots was like "listening to the terrain with a magnifying glass". Finally, Mars's atmosphere muffles different frequencies than Earth's atmosphere. Therefore, additional sound processing is required to perceive Mars' sounds as we do on Earth. Hear an audio Mars-Earth comparison in [27].

## V. CONCLUSION

### A. Effectiveness of sound transparency

The main finding of this study is the **quantification** of a drop in the accuracy in the sense of self-location when sound masking is applied and eyes are closed. This data can be used to argue the case for sound transparency for spacesuits, as visual occlusions of terrain are not rare. Compared to haptics, enhancing sensory input with sound transparency is simpler and lighter. Following we provide a cost-benefit analysis.

### B. Cost-benefit analysis

#### 1) Costs

It is complex to estimate the absolute repercussion of adding a subsystem to a spacesuit. However, it is straightforward to calculate the percentual increase if we assume costs are proportional to the mass [29]. A typical EVA suit weights 115kg. Commercially available noise-canceling cockpit headsets, (similar electronics), weight ~250g. Applying this rule the predicted increase is **0.2%**.

#### 2) Benefits of sound transparency
- Reduction of a trip over during an EVA
- Mitigation of sensory deprivation effect
- Reduction of cognitive load
- Improved self-location during temporary occlusions

#### 3) Analysis

The economic impact due to lower accident risk must be now weighted against a 0.2% increase in cost. However, further tests to quantify these theoretical reductions with **actual** space suits are desirable. Other areas of application of these findings are underwater welding operations as they have similar impairments and record one of the highest occupational fatality rates in the world [28].

## APPENDIX

Link to visualizations, data:
https://www.kaggle.com/harriken/fukuda-unterberger

## ACKNOWLEDGMENT

To H. Vallery, for insightful comments; M. Faris, for 3D printing of prototypes; A. Sheimy, HaptX staff for wearable electronics advice.